\documentstyle[useAMS,epsfig]{mn2e}

\def\l{{\ell}}

\def\healpix{H{\sc ealpix }}
\def\glesp{G{\sc lesp }}

\def\wmap{\hbox{\sl WMAP~}}

\def\etal{et al.~}
\def\c{{\rm cut}}

\def\alm{a_{\l m}}
\def\ylm{Y_{\l m}}

\def\lm{{\l m}}
\def\lmax{\l_{\rm max}}
\def\mmax{m_{\rm max}}

\def\C{{\bf C}}
\def\S{{\bf S}}

\newcommand{\nbi}{{Niels Bohr Institute, Blegdamsvej 17,
DK-2100 Copenhagen, Denmark}}
\newcommand{\asiaa}{{Institute of Astronomy and Astrophysics, Academia Sinica, P.O.Box 23-141, Taipei 10617, Taiwan, R.O.C.}}
\newcommand{\sfu}{{Southern Federal University, Space Research Department,
Zorge,5, 344091, Russia}}

\title[Phase analysis of the cosmic microwave background from an incomplete sky coverage]{Phase analysis of the cosmic microwave background from an incomplete sky coverage}  
\author[Chiang \& Naselsky]{Lung-Yih Chiang$^{1,2}$\thanks{in alphabetic order} and Pavel D. Naselsky$^{2,3}$ \\  \\
$^1$ \asiaa \\
$^2$ \nbi \\ 
$^3$ \sfu}

\date{Accepted 2007 ???? ???; Received 2007 ???? ???}

\begin{document}
\maketitle

\begin{abstract} 
Phases of the spherical harmonic analysis of full-sky cosmic microwave background (CMB) temperature data contain useful information complementary to the ubiquitous angular power spectrum. In this letter we present a new method of phase analysis on incomplete sky maps. It is based on Fourier phases of equal-latitude pixel rings of the map, which are related to the mean angle of the trigonometric moments from the full-sky phases. It has an advantage for probing regions of interest without tapping polluted Galactic plane area, and can localize non-Gaussian features and departure from statistical isotropy in the CMB.

\end{abstract}


\section{Introduction}
The temperature anisotropy of the cosmic microwave background (CMB) radiation contains a wealth of information about our Universe. Its statistical properties not only shed light on what kind of universe we are living in, but also lay the foundation for the significance and interpretation of the angular power spectrum. According to the generally accepted cosmological model, namely the Cosmological Concordance Model, the primordial fluctuations in the early Universe constitute a Gaussian random field (GRF) \cite{bbks,be}. As the CMB is an observable imprint of the primordial fluctuations, therefore, after the NASA \wmap data release \cite{wmapresults,wmapfg,wmapcl,wmapng,wmap3ytem,wmap3ycos}, testing the Gaussianity of the CMB has been imperative for our understanding of the Universe  \cite{wmaptacng,gaztanagz,coleskuiper,park,eriksenmf,santanderng,romanng,hansen,mukherjee,larson,phaserandomwalk,edingburgh,mnn,toh,dtzh,eriksenasym,copi,schwarz,evil,bernui,abramolow,wmap3yrng,cruz,mcewenwmap3,copi3y,fgxcorr,eriksen3yasym}.  

One of the most general ways to test Gaussianity is based on the ``random phase hypothesis'', as any departure from Gaussianity in the data shall register as some sort of phase correlation in the harmonic domain. There have been several non-Gaussianity methods devised from phase information: Shannon entropy of phases  \cite{phaseentropy}, phase mapping \cite{wmaptacng}, trigonometric moments \cite{ndv04}, phase sums  \cite{matsubara,hikage}, random walks \cite{stannardrandomwalk,phaserandomwalk}, some of which have been deployed on \wmap full-sky maps and detection of non-Gaussianity has been decleared.

As phases and morphology are closely related \cite{morph}, one requirement for applying phases as a useful statistical diagnostic is continuity of the boundaries in the data, otherwise the phases would faithfully reflect boundary discontinuity by strong coupling. Therefore, those above-mentioned methods using phase information (particularly for CMB studies) can be deployed only on data with a full-sky coverage. 

Due to excessive foreground contamination near the Galactic plane, the \wmap science team has adopted a specific foreground removal strategy using the so-called temperature masks \cite{wmapfg,wmap3ytem}, which divide the full sky into 12 regions. The largest, {\it Region 0}, covers about 89\% of the full sky, whereas the other 11 regions are masked due to heavy foreground emissions of different kinds around the Galactic plane: synchrotron, free-free and dust emission (see Fig.1). Although a full-sky derived CMB map, the Internal Linear Combination (ILC) map, is combined from the 12 foreground-reduced regions and available to the public, most scientific results including the angular power spectrum are derived from the cleanest {\it Region 0} \cite{wmap3ytem}, and the full-sky ILC map is known to still have foreground residuals near the Galactic plane.

In this letter we present a new method for phase analysis on maps with Galaxy cut, assuming that the orthogonality of the Fourier series in the azimuthal direction outside the Galaxy cut is still preserved\footnote{Note that \wmap {\it Region 0} is not symmetric with respect to $b=0$, but $|b| > 30^\circ$ is surely outside the Galaxy mask (see Fig.1).}. It is based on Fourier phases of equal-latitude pixel rings of the map, which is closely related to the mean angle of the trigonometric moments on the full-sky phases with some weighting coefficients \cite{phaserandomwalk}. We can examine the Fourier phases of all equal-latitude pixel rings from regions, e.g. \wmap {\it Region 0}, while avoiding the polluted Galactic plane area. More importantly, we can pin down non-Gaussian features by using the phases derived this way, an advantage that is generally lacking in the analysis processed in harmonic domain. Note that all the above mentioned methods based on phases can be applied using the phases we derive in this letter.  

\begin{figure}
\begin{center}
\epsfig{file=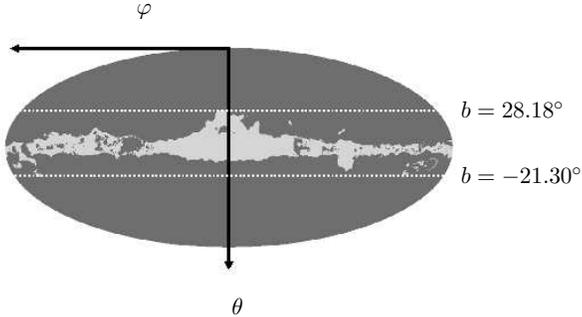,width=14cm}
\end{center}
\caption{The \wmap masks with the polar coordinates $(\theta, \varphi)$. The dark gray region is the \wmap {\it Region 0}, covering 89\% of the full sky. The light gray area covers the \wmap {\it Region 1} to {\it 11} (the Galaxy mask). The white dotted lines denote the Galactic latitude $b=28.18^\circ$ ($\theta=1.079$) and $-21.30^\circ$ ($\theta=1.943$), the boundary of the Galaxy mask.}
\end{figure}

\section{Phases from an incomplete sky map}
The standard treatment for a full-sky CMB signal $T(\theta,\varphi)$ is via spherical harmonic decomposition:
\begin{eqnarray}
T(\theta,\varphi)=\sum_{\l=0}^{\lmax} \sum_{m=-\l}^{\l} \alm Y_{\lm}(\theta,\varphi),
\label{eq1}
\end{eqnarray}
where $\lmax$ is the maximum multipole number used in map, $\theta$ and $\varphi$ are the polar and azimuthal angle, respectively, and $\alm$ are the spherical harmonic coefficients. $\ylm$ are the spherical harmonics, defined in terms of Legendre Polynomials:
\begin{equation}
\ylm(\theta,\varphi)=N_\lm P^m_\l(\cos \theta) \exp(i m \varphi),
\label{sh}
\end{equation}
where 
\begin{equation}
N_\lm=(-1)^m \sqrt{\frac{(2\l+1)(\l-m)!}{4\pi(\l+m)!}}.
\end{equation}
The $\alm$ coefficients can be further written as $\alm=|\alm| \exp[i \Phi_{\lm}]$, where $\Phi_{\lm}$ are the phases. If the CMB temperature  anisotropies constitute a GRF, the real and imaginary part of the $\alm$ are both Gaussian distributed, or equivalently, the $|\alm|$ are Rayleigh distributed and phases $\Phi_\lm$ are uniformly random in $[0, 2\pi]$. In polar coordinate system $\theta=\pi/2$ is associated with the Galactic plane ($b=0$), as used by \healpix \cite{healpix} and \glesp \cite{glesp} software packages.

For signal from an incomplete sky coverage, implementation of the spherical harmonic decomposition is no longer correct, as the orthogonality of the spherical harmonics $\ylm$ is broken \cite{gorskiin}. This is particularly the case when one is to analyze the \wmap ILC Galaxy-cut map. Nevertheless, Galaxy cut only breaks the orthogonality of the spherical harmonics over $\theta$ direction, but not $\varphi$ outside Galaxy cut \cite{gorskiin}.

To see how phases of an incomplete sky map (e.g. ILC Galaxy-cut map) can be related to its full-sky phases, let us extract an equal-latitude pixel ring at  $\theta=\theta_c$, where $\theta_c$ is outside the maximum latitude of any Galaxy masks. This ring $ T(\theta_c,\varphi)\equiv T_c(\varphi)$ is now one-dimensional signal, for which we can use a Fourier Transform approach with coefficients $g^c_m$: 
\begin{equation} 
T_c(\varphi)=\sum_{m=-\mmax}^{\mmax} g^c_m \,\exp(i m \varphi),
\label{eq2}
\end{equation}
where $\mmax\le \lmax$ and 
\begin{equation}
g^c_m=\frac{1}{2\pi}\int_0^{2\pi} d\varphi \, T_c(\varphi)\, \exp(-im\varphi).
\end{equation}
We can then relate the ring with the full-sky signal via Eq.(\ref{eq1}) and (\ref{sh}) and get 
\begin{equation}
g^c_m=\sum_{\l\ge |m|}^{\lmax} N_\lm\, P_\l^m(\cos\theta_c)\, \alm.
\label{eq3}
\end{equation}
That is, the Fourier coefficients $g^c_m$ of the ring can be expressed as a combination of the full-sky $\alm$. Writing $g^c_m = |g^c_m|\exp(i \kappa^c_m)$, the phases $\kappa^c_m$ are
\begin{equation}
\kappa^c_m=\tan^{-1}\frac{\sum_{\l\ge |m|}^{\lmax} W_\lm(\theta_c) \sin\Phi_\lm}
{\sum_{\l \ge |m|}^{\lmax} W_\lm(\theta_c) \cos \Phi_\lm},
\label{eq4}
\end{equation}
where $W_\lm (\theta_c) = N_\lm P_\l^m(\cos\theta_c)|\alm|$.
Note that the phases $\kappa_m$ correspond to the ``mean angle'' of all $\Phi_{\lm}$ with some weighting coefficients $W_\lm(\theta_c)$ involving the $|\alm|$ \cite{phaserandomwalk}. If the ring $T(\theta_c,\phi)$ is taken from a GRF, its phases $\kappa^c_m$ are a combination of the uniformly random phases $\Phi_\lm$, hence are also uniformly random in $[0, 2\pi]$. We can then examine all the pixel rings of the ILC map for $0\le  \theta \le \pi/3$ and $2\pi/3\le  \theta \le\pi$ without tapping the heavily polluted region near the Galactic plane. Our demonstration here is a special case for a well known theory: any $N-n$ dimensional cross sections of $N$ dimensional Gaussian random process produce a Gaussian process as well. Thus, if one is to investigate the phases of the $\alm$ coefficients from a full-sky map, one can test alternatively the phases of equal-latitude pixel rings of the Galactic-cut map.

\begin{figure}
\epsfig{file=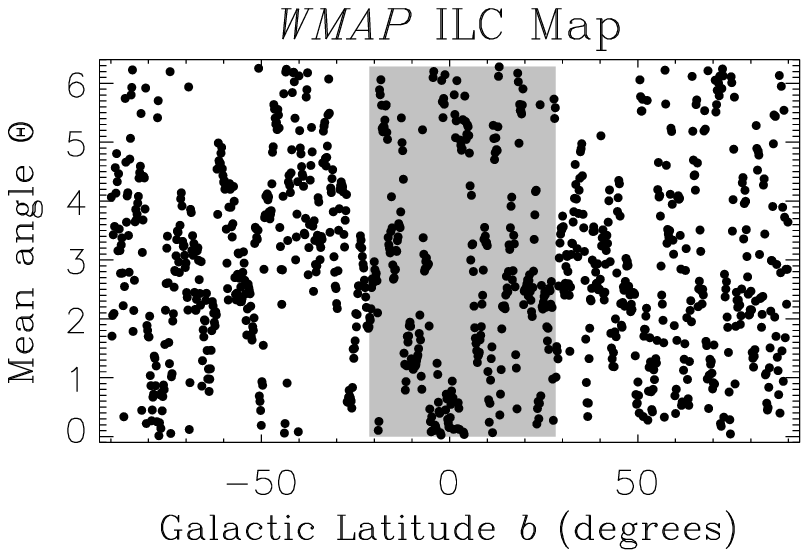,width=9cm}
\epsfig{file=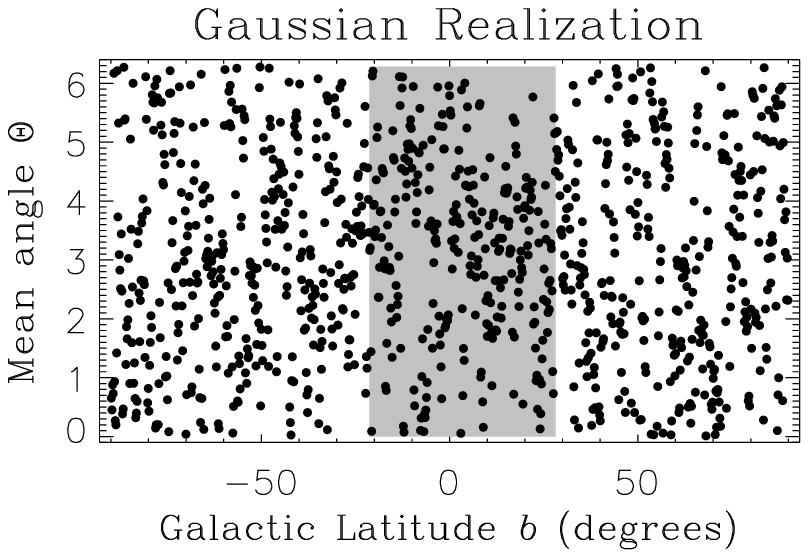,width=9cm}
\caption{The mean angle (defined in Eq.(\ref{trigo}) and (\ref{ma}) with $\Delta m=1$ up to $M=50$) of the Fourier phases from equal-latitude pixel rings $T_c(\varphi)$ of the \wmap ILC 3-year map (top) and of a Gaussian realization (bottom). The gray area denotes the Galatic latitude boundary of the \wmap Galaxy mask at $[-21.30^\circ,28.18^\circ]$ (see Fig.1). One can see that the mean angles of the ILC map are fairly non-random, compared with the Gaussian realization.}\label{meanangle50}
\end{figure}

\begin{figure}
\epsfig{file=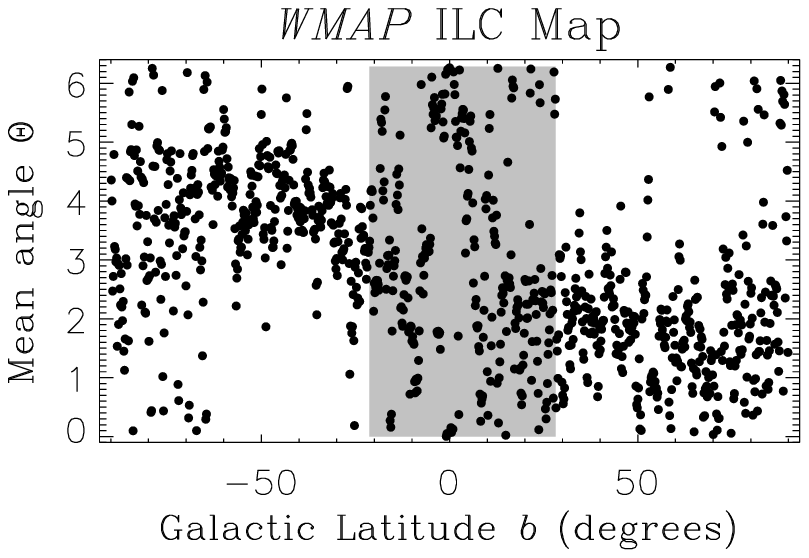,width=9cm}
\epsfig{file=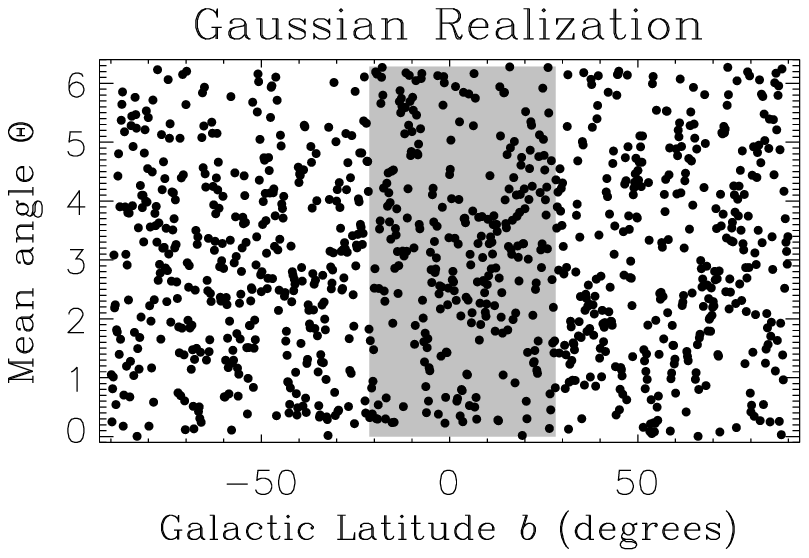,width=9cm}
\caption{The mean angle (defined in Eq.(\ref{trigo}) and (\ref{ma}) with $\Delta m=1$ up to $M=300$) of the Fourier phases from equal-latitude pixel rings $T_c(\varphi)$ of the \wmap ILC 3-year map (top) and of a Gaussian realization (bottom). The gray area denotes the Galactic latitude boundary of the \wmap Galaxy mask at $[-21.30^\circ,28.18^\circ]$ (see Fig.1). One can see that the mean angles of the ILC map are significantly non-random, compared with the Gaussian realization.}\label{meanangle300}
\end{figure}

However, a more intriguing question is whether we can reconstruct the phases of a full-sky signal $\Phi_\lm$ by using the phases $\kappa_m$ from the stripes of an incomplete sky map? Obviously we cannot reconstruct all the phases due to Galaxy cut, but we can recover significant part of the full-sky phases. Based on G\'{o}rski (1994) method and taking into account that Galaxy cut map only breaks the orthogonality of the Legendre polynomials in $\theta$ direction, there shall exist some polynomials $K_\l^m(\theta)$ which are orthogonal to the Legendre polynomials $P_i^m(\theta)$ within some intervals $[0,\pi/2-\theta_\c]$ and $[\pi/2+\theta_\c,\pi]$. Namely,
\begin{eqnarray}
\int_{x_\c=\cos \theta_\c}^1 dx P_\l^m(x) K_{\l'}^m (x)=F(\l,m)\delta_{\l \l'},
\label{eq5}
\end{eqnarray}
where $F(\l,m)$ is the normalization coefficient.
Then, defining new coefficients
\begin{eqnarray}
S^{+}_\lm&=& \int_{x_\c}^1 dx g_m(x) K_\l^m (x)  \\ \nonumber
&=&  N_{\lm} F(\l,m) |\alm| \exp(i \Phi_\lm); \\ \nonumber
S^{-}_\lm&=&  (-1)^m \int_{-1}^{-x_\c} dx g_m(x) K_\l^m (x) \\ \nonumber 
&=& N_{\lm} F(\l,m) |\alm| \exp(i \Phi_\lm),
\label{eq6}
\end{eqnarray}
which we can use for analysis of their phases. Since $F(\l,m)$ is a sign-flipping function, the phases of $S^{+}_\lm$ are equivalent to $\Phi_\lm \pm \pi$. However, the cross correlation of phases can be preserved. Care has to be taken in deconvolution for the phases. Due to pixelization of the signal, particularly for the polar caps, modes at high multipole numbers tap the window function of the pixels. Implementing simple deconvolution of the signal by window functions produces artifacts, which needs to be corrected by Tikhonov regularization. The same correction is needed for the high $m$ modes as they are close to the Nyquist frequency. We will describe this approach in another paper.

\section{Mean angle of the phases from the ILC (Galaxy-cut) map}
In this section, serving as an example of the Fourier phases $\kappa_m$ providing a useful diagnostic, we employ the trigonometric moments and the mean angles on the phases derived from the equal-latitude pixel rings. The trigonometric moments are defined as follows \cite{phaserandomwalk}:
\begin{eqnarray}
\C_c( \Delta m)&=&\sum_{m=1}^{M} \cos\left(\kappa^c_{m+\Delta m}-\kappa^c_m\right); \nonumber \\
\S_c( \Delta m)&=&\sum_{m=1}^{M} \sin\left(\kappa^c_{m+\Delta m}-\kappa^c_m\right),
\label{trigo}
\end{eqnarray}
where $M \le \lmax -\Delta m$. Note that in this definition we use phase differences where $\Delta m \ge 1$. The mean angle is defined as  
\begin{equation}
\Theta_c(\Delta m)=\tan^{-1}\frac{\S_c(\Delta m)}{\C_c(\Delta m)}.
 \label{ma}
\end{equation}
The mean angle can be seen as the resultant angle of Pearson's random walk (walks with a fix length in each step) : $\sum^M \exp[i(\kappa^c_{m+\Delta m}-\kappa^c_m)]$ \cite{pearson,phaserandomwalk}.
For a GRF, the phases $\Phi_\lm$ are uniformly random, so are the $\kappa_m$ for each pixel ring. As the difference of any two random variables should be random as well, one then expects the mean angles $\Theta$ from an ensemble of Gaussian processes to be uniformly random in $[0, 2\pi]$.

We use the \wmap ILC 3-year map with $\lmax=512$ as an example of a high-resolution map. For each equal-latitude pixel ring $T_c(\varphi)$, we use Fast Fourier Transform and obtain the phases $\kappa^c_m$. In Fig.\ref{meanangle50} and \ref{meanangle300} we plot the mean angles of each pixel rings with $\Delta m=1$ up to $M=50$ and 300 respectively against the Galactic latitude $b$. In each figure, the top panel is the mean angles for ILC pixel rings and bottom for those from a Gaussian realization with the \wmap best-fit $\Lambda$CDM power spectrum. The gray area denotes the Galactic latitude boundary of the \wmap Galaxy mask at $[-21.30^\circ,28.18^\circ]$ (see Fig.1). From both Fig.\ref{meanangle50} and  \ref{meanangle300}, one can see the ILC map outside the Galaxy mask has significant non-uniform distribution for the mean angles $\Theta$ whereas for the Gaussian realization $\Theta$ are fairly uniformly random. Note that this example is for illustration purpose only, and more thorough analysis will be present in another paper.


\section{Conclusion}
In this Letter we have presented a new method of phase analysis of the CMB from an incomplete sky coverage. It is based on Fourier phases of equal-latitude pixel rings of the underlying map, which are, theoretically speaking, related to the mean angles of full-sky phases via well-defined weighting coefficients. We have also employed trigonometric moments and mean angles on the new phases, which has shown qualitatively significant non-random distribution of the mean angles, signature of departure of Gaussianity. We would like to emphasize that all the methods developed on using the full-sky phases can be easily implemented on the phases from an incomplete sky coverage. We will examine in details of non-Gaussianity using these phases in the next paper. 

\section*{Acknowledgments}

We acknowledge the use of the Legacy Archive for Microwave Background
Data Analysis (LAMBDA). We also acknowledge the use of \healpix package \cite{healpix} to
produce $\alm$. The \glesp package \cite{glesp} was used in this work.

\newcommand{\autetal}[2]{{#1\ #2. \etal,}}
\newcommand{\aut}[2]{{#1\ #2.,}}
\newcommand{\saut}[2]{{#1\ #2.,}}
\newcommand{\laut}[2]{{#1\ #2.,}}

%
%
\newcommand{\refs}[6]{#5, #2, #3, #4} 
\newcommand{\unrefs}[6]{#5, #2 #3 #4 (#6)}  

%
%
\newcommand{\book}[6]{#5, {\it #1}, #2} 
%
\newcommand{\proceeding}[6]{#5, in #3, #4, #2} 
%
%

\newcommand{\combib}[3]{\bibitem[\protect\citename{#1 }#2]{#3}} 

%
%
\def\apj{ApJ}
\def\apjl{ApJL}
\def\apjs{ApJS}
\def\mn{MNRAS}
\def\nature{nat}
\def\aa{A\&A}
\def\prl{Phys.\ Rev.\ Lett.}
\def\prd{Phys.\ Rev.\ D}
\def\pr{Phys.\ Rep.}
\def\ijmpd{Int. J. Mod. Phys. D}
\def\jcap{J. Cosmo. Astro. Par.}

\def\cup{Cambridge University Press, Cambridge, UK}
\def\princetonpress{Princeton University Press}
\def\worldpress{World Scientific, Singapore}
\def\oxfordpress{Oxford University Press}

\end{document}